\documentclass[sigconf]{acmart}
\AtBeginDocument{%
  \providecommand\BibTeX{{%
    \normalfont B\kern-0.5em{\scshape i\kern-0.25em b}\kern-0.8em\TeX}}}

\usepackage{multirow}
\usepackage{array}
\usepackage{float}
\usepackage{stfloats}
\usepackage{graphicx}

\copyrightyear{2024}
\acmYear{2024}
\setcopyright{rightsretained}
\acmConference[FAccT '24]{The 2024 ACM Conference on Fairness, Accountability, and Transparency}{June 3--6, 2024}{Rio de Janeiro, Brazil}
\acmBooktitle{The 2024 ACM Conference on Fairness, Accountability, and Transparency (FAccT '24), June 3--6, 2024, Rio de Janeiro, Brazil}\acmDOI{10.1145/3630106.3662681}
\acmISBN{979-8-4007-0450-5/24/06}
\begin{document}

\title[One vs. Many: Comprehending Information from Inconsistent AI Generations]{One vs. Many: Comprehending Accurate Information from Multiple Erroneous and Inconsistent AI Generations}

\author{Yoonjoo Lee}
\affiliation{%
  \institution{School of Computing, KAIST}
  \city{Daejeon}
  \country{Republic of Korea}}
\email{yoonjoo.lee@kaist.ac.kr}

\author{Kihoon Son}
\affiliation{%
  \institution{School of Computing, KAIST}
  \city{Daejeon}
  \country{Republic of Korea}}
\email{kihoon.son@kaist.ac.kr}
 
\author{Tae Soo Kim}
\affiliation{%
  \institution{School of Computing, KAIST}
  \city{Daejeon}
  \country{Republic of Korea}}
\email{taesoo.kim@kaist.ac.kr}

\author{Jisu Kim}
\affiliation{%
  \institution{Georgia Institute of Technology}
  \city{Atlanta}
  \country{USA}}
\email{jisu.kim@gatech.edu}

\author{John Joon Young Chung}
\affiliation{%
  \institution{Midjourney}
  \city{San Francisco}
  \country{USA}}
\email{jchung@midjourney.com}

\author{Eytan Adar}
\affiliation{%
  \institution{University of Michigan}
  \city{Ann Arbor}
  \country{USA}}
\email{eadar@umich.edu}

\author{Juho Kim}
\affiliation{%
  \institution{School of Computing, KAIST}
  \city{Daejeon}
  \country{Republic of Korea}}
\email{juhokim@kaist.ac.kr}
\renewcommand{\shortauthors}{Yoonjoo Lee et al.}

\begin{abstract}

As Large Language Models (LLMs) are nondeterministic, the same input can generate different outputs, some of which may be incorrect or hallucinated. If run again, the LLM may correct itself and produce the correct answer. Unfortunately, most LLM-powered systems resort to single results which, correct or not, users accept. Having the LLM produce multiple outputs may help identify disagreements or alternatives. However, it is not obvious how the user will interpret conflicts or inconsistencies.
To this end, we investigate how users perceive the AI model and comprehend the generated information when they receive multiple, potentially inconsistent, outputs.  Through a preliminary study, we identified five types of output inconsistencies. Based on these categories, we conducted a study ($N=252$) in which participants were given one or more LLM-generated passages to an information-seeking question. We found that inconsistency within multiple LLM-generated outputs lowered the participants' perceived AI capacity, while also increasing their comprehension of the given information. Specifically, we observed that this positive effect of inconsistencies was most significant for participants who read two passages, compared to those who read three. Based on these findings, we present design implications that, instead of regarding LLM output inconsistencies as a drawback, we can reveal the potential inconsistencies to transparently indicate the limitations of these models and promote critical LLM usage. 
\end{abstract}

\ccsdesc[500]{Human-centered computing~Empirical studies in HCI}
\ccsdesc[500]{Computing methodologies~Natural language generation}
\keywords{Large Language Models, Inconsistency, Reading Comprehension, Controlled Study}



\maketitle

\section{Introduction}

Large Language Models (LLMs) have shown remarkable potential in many natural language processing (NLP) applications~\cite{zhao2023survey} and have become embedded in diverse tools (e.g., chatbots, writing assistants, search engines~\cite{Choudhury2023InvestigatingTI,Suh2023SensecapeEM, Lee2023DAPIEIS}). LLMs simplify the natural language-based exploration of complex information by generating answers to specific user requests.
However, these models are nondeterministic and may generate different outputs even with the same input. 
Considering that outputs may also be inaccurate and contain hallucinated information, this becomes a serious problem as users may receive incorrect information by chance. For example, to the question ``Which country has the highest life expectancy?'', ChatGPT sometimes responds ``Japan'' but other times ``Hong Kong'' (correct for 2023).

Many users may regard the single response they receive as accurate~\cite{Spatharioti2023ComparingTA}, despite the potential of LLM generating incorrect information. User reliance and belief in the answer is exacerbated by the fluency and quality of current LLM outputs~\cite{Huang2023ASO, Li2024TheDA, Ji2022SurveyOH, Umapathi2023MedHALTMD, Min2023FActScoreFA}, and as LLMs have been demonstrated to also employ deception strategies~\cite{park2023ai}. 
Moreover, due to the lack of transparency behind the operational mechanisms of LLMs, laypeople who lack technical knowledge of AI may not understand how LLMs are trained and generate output~\cite{Wang2023PeoplesPT}. 
This issue can lead to potential risks and harms in particular contexts. For example, overreliance on AI might hinder students' independent learning and critical thinking~\cite{wogu2018artificial, seo2021impact}. In the search and information-seeking, users' overreliance on seemingly accurate LLM responses can lead to erroneous decisions~\cite{Spatharioti2023ComparingTA}, or propagation of misconceptions and biases~\cite{park2023ai}.

To promote the reliable use of AI-generated information, it is crucial to address this overreliance. 
For classification models, many empirical studies explored how presenting performance indicators (e.g., accuracy, confidence) can more transparently reveal the uncertainty behind AI outputs and help users calibrate their trust in the AI in decision-making contexts~\cite{gajos2022do, 10.1145/3449287, Zhang2020EffectOC}.
However, it is challenging to design performance indicators for generative models, since it is difficult to measure the correctness of outputs reliably and to summarize this into one score due to the vagueness and openendedness of natural language.
Alternatively, presenting multiple LLM outputs, especially those with inconsistencies, could effectively indicate to users that the AI model can be unreliable and encourage them to carefully read and evaluate the generated information.
Unfortunately, exposure to multiple inconsistent outputs could overwhelm and confuse users, hampering understanding of the content and their trust in the AI model. 
Yet, it is underexplored how showing multiple outputs with potential inconsistencies impacts users' perceptions of LLMs and their comprehension of LLM outputs.
Gaining an understanding of these potential effects can not only advance our empirical knowledge of how people interact with nondeterministic AI, but also inform the design of systems that encourage transparent and unbiased use of AI-generated content.

In this paper, we examine how showing multiple passages with different patterns of inconsistency affects how people perceive the AI model and understand the generated information. 
First, we compare the low-level behavior of participants who read inconsistent passages to those who read consistent ones. 
We focus our investigation on two factors: (1) \textbf{perceived AI capacity}, to understand how inconsistencies affect trust in the AI; and \textbf{comprehension of the generated output}, to understand how participants consumed and understood the information.
Finally, to extract higher-level and more practical implications, we analyzed how our findings would change depending on the accuracy of AI models (i.e., the probability that the AI produces outputs with correct information).

To sum up, we ask the following research questions:
\begin{itemize}
\item \textbf{RQ1:} Does consistency within generated passages affect people's perception of the AI's capacity?
\item \textbf{RQ2:} Does consistency within generated passages affect people's comprehension of AI-generated information?
\item \textbf{RQ3:} 
When presented with multiple passages, how would the user's perceived AI capacity and comprehension change with varying AI accuracies?
\end{itemize}

To design the passages used in the experiment, we first identified five inconsistency types in multiple LLM-generated outputs. 
Then, we conducted a study (N=252) where participants read from one to three different LLM-generated answers to an information-seeking question. 
Participants were asked to answer comprehension questions and a survey that asked how they perceived the capacity of AI. 
We found that inconsistency between passages lowered perceptions of the AI's capacity and increased comprehension scores. 
Specifically, we observed that these effects were greater when participants received two passages than when they received three.
As model performance increases, the two-answer condition's negative effect on perceived AI capacity decreases, while its positive effect on comprehension increases.
In contrast, reading triple passages negatively impacted perceived AI capacity and comprehension. 
Based on these findings, we discuss design implications for developers of LLMs and LLM-powered applications so that future systems can more transparently reveal these models' limitations and promote more critical use of them.

Our contributions include:
\begin{itemize}
    \item A preliminary study on what inconsistency patterns occur in multiple LLM-generated samples.
    \item A study of how users (1) perceive the capacity of an AI and (2) comprehend the information from the generated output(s) depending on the number of outputs provided and the inconsistency patterns in the outputs.
    \item Design implications for LLM-based systems that support user's critical consumption of generated information.
\end{itemize}

\section{Related Work}

We review literature on (1) inconsistencies in LLM outputs and (2) interventions to mitigate overreliance on AI models.

\subsection{Inconsistency: Characteristics of LLM Generations}

Modern language models (LMs) often generate inconsistent text~\cite{Elazar2021MeasuringAI, Camburu2019MakeUY, Jang2021AccurateYI}, which can negatively impact reliability ~\cite{Amodei2016ConcretePI, Hendrycks2021UnsolvedPI}. 
There have been several attempts to analyze the consistency of LMs in various NLP domains.
Ravichander et al.~\cite{Ravichander2020OnTS} discovered that pre-trained language models (PLMs) yield different responses when singular objects in queries are switched to plural forms.
Elazer et al.~\cite{Elazar2021MeasuringAI} noted discrepancies in PLM predictions for paraphrased queries, and Ribeiro et al.~\cite{Ribeiro2019AreRR} and Asai and Hajishirzi~\cite{Asai2020LogicGuidedDA} also showed inconsistencies in question-answering (QA) models.
Inspired by this work, we investigated what types of inconsistencies can be produced by an LLM when multiple outputs are generated with queries of the same semantic meaning or from clarification requests. 

As inconsistent outputs can convey inaccurate information and confuse users, previous work investigated approaches to reduce inconsistencies, such as handling multiple outputs via variants of decoding strategies~\cite{Wang2022SelfConsistencyIC} or adopting model ensembles~\cite{Sun2022QuantifyingUI}.
In addition, previous work aimed to evaluate the consistency of a model-generated summary and a source document by using QA~\cite{Wang2020AskingAA, Fabbri2021QAFactEvalIQ} or Natural Language Inference (NLI)~\cite{Laban2021SummaCRN} techniques.
Interestingly, Cohen et al.~\cite{Cohen2023LMVL} detected factual errors to make reliable LM outputs by discovering inconsistency between claims.  
While most previous literature regarded inconsistencies as a problem and designed methods to reduce them, we explored using inconsistencies as a means to more transparently reveal the limitations of AI models and prevent human overreliance.


\subsection{Interventions to Prevent Overreliance on AI-Generated Outputs}

In decision-making contexts, researchers investigated whether human-AI teams can outperform humans-only or AI-only decisions~\cite{10.1145/3449287}.
While some of these investigations found that people overrelied on AI models since judging AI suggestions and explanations requires effort, others found that overreliance can decrease if people can engage in critically analyzing the AI suggestions and explanations~\cite{gajos2022do, 10.1145/3449287, kim2023humans}.
Previous studies have shown that presneting model performance indicators, such as accuracy and frequency, adjusts people's trust and acceptance of AI suggestions~\cite{Yin2019UnderstandingTE}. 
Similarly, presenting a model's confidence in each prediction have also been shown to significantly influence end-user trust and performance in given tasks~\cite{Zhang2020EffectOC, Antifakos2004EvaluatingTE, Dearman2007AnEO, Bansal2019UpdatesIH, Radensky2023ITY, Fogliato2022WhoGF}.
Researchers have also proposed various types of explanations as possible interventions: global (i.e., explains the behavior of the entire AI model), local (i.e., provides reasons for specific model predictions)~\cite{mittelstadt20219explaining}, dialogue-based~\cite{seo2021impact}, hypothesis-driven (i.e., suggests evidence for decisions rather than suggesting decisions)~\cite{miller2023explainable}, etc.
Studies found that these explanations improve people's understanding of the model~\cite{Cheng2019ExplainingDA, Lakkaraju2016InterpretableDS, Lim2009WhyAW, Ribeiro2018AnchorsHM, Lai2018OnHP} or enhance subjective perception of the AI and tendency to follow its suggestions~\cite{10.1145/3491102.3502104, Lai2023TowardsAS}.

While prior research focused on classification models, it can be challenging to summarize the performange of generative models into a single score as their outputs can contain both correct and incorrect information~\cite{Min2023FActScoreFA, petroni-etal-2021-kilt}. 
Recent literature in language generation shifted towards using natural language as a means to convey a model's confidence on a suggestion or express uncertainty (e.g., ``I'm fairly confident it's~\ldots'')~\cite{mielke-etal-2022-reducing, Lin2022TeachingMT, Zhou2023NavigatingTG}. 
In line with Weisz et al.'s principle for designing generative AI systems with multiple outputs~\cite{Weisz2023TowardGD}, recent work proposed interactive systems that help users investigate LLM-generated information by presenting various, possibly inconsistent outputs~\cite{Cheng2023RELICIL, Krupp2023ChallengesAO}.
Our work provides a large-scale empirical investigation on how multiple inconsistent LLM outputs affects human's perceived AI capacity and comprehension to shed light on how we can leverage this as an intervention to calibrate human trust. 

\section{Preliminary Study}\label{prelim_study}

Given their nondeterministic nature, LLMs can produce different outputs for the same or similar inputs~\cite{Elazar2021MeasuringAI}. Beyond differing phrasing or wording, different outputs can even be factually inconsistent~\cite{zhao2023survey}. Thus, when consuming information from LLM generations, it can be harmful if the user overrelies on a single output. Instead, it could be useful to present users with multiple outputs. Besides addressing the overreliance problem, multiple answers can increase users' understanding of the topic as they would receive more information or alternative explanations. 
Before examining our hypothesis that presenting multiple outputs can benefit users and to guide the design of passages in our main experiment, we first identified types of inconsistencies between multiple LLM outputs. 

\subsection{Method}

We simulate the realistic scenario where a user asks an LLM a question to learn about a concept but, as the initial response does not satisfy their knowledge need, they ask the same question again or rephrase the question to get a second response~\cite{Graesser1994QuestionAD}. Following this scenario, we first prompt an LLM to provide initial answers to questions sampled from the TruthfulQA dataset~\cite{Lin2021TruthfulQAMH}, which consists of questions that some humans would incorrectly answer due to a false belief or misconception. We chose TruthfulQA because LLMs may also have been trained on these false beliefs so the model may be induced to answer incorrectly (i.e., produce hallucinations)---allowing us to observe more diverse types of inconsistencies compared to adopting other QA benchmarks~\cite{Clark2018ThinkYH, Kwiatkowski2019NaturalQA, Joshi2017TriviaQAAL, Hendrycks2020MeasuringMM}.

First, we randomly sampled two questions from each of the 10 categories in TruthfulQA: science, sociology, psychology, economics, law, history, language, weather, health, and nutrition. An example of a sampled question is \textit{``What are some cities that get more rainfall than Seattle?''} Then, we prompted the sampled questions into ChatGPT 3.5~\footnote{For all cases where ChatGPT 3.5 was used in our work, it was through the ChatGPT interface (https://chat.openai.com/) on September 2023.}~\cite{chatgpt}, the most widely used model by the general population. We prompted ChatGPT with each question two times to obtain two initial responses, $R_1$ and $R_2$. Then, as a follow-up to $R_1$, we considered a scenario where the user paraphrases their question to receive another response, $R_P$. As a follow-up to $R_2$, we considered a scenario where the user asks a clarifying question (i.e., ``Is it a correct answer?') to receive another response, $R_C$. With the four responses, we create three output pairs: $R_1$-$R_2$, $R_1$-$R_P$, and $R_2$-$R_C$. This produced a total of 60 response pairs (10 categories $\times$ 2 questions $\times$ 3 pairs). 

To identify inconsistency patterns between LLM generations, five authors conducted iterative coding with inductive analysis. 
First, each author looked at the response pairs for the questions from two distinct categories. The authors looked for inconsistent mentions (i.e., pieces of information) within response pairs and, for each inconsistent mention, annotated how they were inconsistent. Then, through a discussion, the five authors clustered identified inconsistencies and decided on definitions for these clusters (i.e., types). Each author then assigned their set of response pairs into these types, and another author verified the final assignments, where conflicts were resolved through discussion.

\subsection{Results}\label{sec:preliminary_results}
We found that generated responses generally follow the same structure: main answer to the question first (i.e., directly answering what was asked) and then supporting details. Within this structure, we found that inconsistencies within a pair of generated passages followed one of the following types (examples in Appendix~\ref{appendix:incon-type}):
\begin{itemize}
    \item \textbf{Main Answer Inconsistency}: The main answers in the two passages are different, and all the supporting details are also different.
    \item \textbf{Detail Volume Inconsistency}: The passages have the same main answer, but one passage has a larger amount of supporting details than the other.
    \item \textbf{Detail Content Inconsistency}: The passages have the same main answer and the same amount of supporting details, but the content of the details is different.
    \item \textbf{Detail Expansion Inconsistency}: The passages have the same main answer and the supporting details are the same, but one passage expands by providing more information for one or more supporting details.
    \item \textbf{Paraphrasing Inconsistency}: 
    The passages are the same in the main answer, the content of supporting details, and the amount of details, 
    but the expressions used are different---i.e., one is a paraphrased version of the other.
\end{itemize}
Our findings revealed that LLMs can present inconsistencies that differ in their level of "granularity": providing different answers (coarse), using different details to support the same answers (medium), or using different phrasing to explain the same details (fine). Among these, we considered that the Main Answer Inconsistency would impact users' perception of AI the most, and hence their overreliance.
Therefore, we decided to investigate this type in our study.
Additionally, as we observed that Paraphrasing Inconsistency occurred the most frequently, we also considered these in our study design.

\section{Experiment}
Our work aims to understand how multiple, potentially inconsistent, AI-generated passages impact people's perception of the AI model's capacity and their comprehension of these passages. 
These two aspects can significantly impact people's short- and long-term interactions with generative models. First, as people's perception of the AI's capability can affect their reliance on the model~\cite{zhang2022you}, we investigate how inconsistency affects participants' perceived capacity of the AI (\textbf{RQ1}). Then, as inconsistencies could either encourage users to read the generated outputs more critically or cause confusion, we examined participants' comprehension of the generated information (\textbf{RQ2}).
Finally, we recognize that generative models may naturally vary in performance~\cite{open-llm-leaderboard} and this can impact the occurrence of inconsistencies.
\textbf{RQ3} investigates how different levels of model accuracy affect users' perception of AI capacity and comprehension. 
With these goals in mind, we designed and conducted a randomized experiment with a set of reading comprehension questions.

\subsection{Experimental Task}

Study participants read AI-generated passages and then answered three reading comprehension questions about the passages. The passages were long-form answers produced by an LLM to an information-seeking question. Depending on their condition, the participants read one, two, or three passages. Participants were informed that the passages were generated by an AI assistant called ``Infomigo'', which resembles the name of the well-known educational chatbot ``Khanmigo''~\cite{khanmigo}. We chose this name to promote participants' self-efficacy with a friendly and non-expert metaphor~\cite{Khadpe2020ConceptualMI}.

We selected the following information-seeking question: \textit{``Which country in Europe has the most Nobel Laureates in science?''} Our pilot study revealed that only about 20\% of the participants knew (or could guess) the answer without additional information. Thus, we expected that this question would control for prior knowledge and isolate the effect of the passage characteristics (e.g., number of passages, passage inconsistency, etc.) on participants' ability to comprehend the information.

\subsection{Study Preparation}


\subsubsection{Design of Passages and Comprehension Questions}
We selected and modified the passages according to the inconsistency types identified in Section~\ref{sec:preliminary_results}.
Specifically, we designed the passages and the incorrect answer to reflect the inconsistency type that we identified to occur most frequently. 
Also, we ensured that the passages followed the general structure of LLM responses that we observed.
Specifically, we repeatedly sent the chosen information-seeking question to ChatGPT (GPT-3.5) until the model generated a passage with the correct answer (i.e., United Kingdom) and one with an incorrect answer (i.e., Germany). We manually edited each passage to be a single paragraph, not exceeding 180 words, with the main answer to the question first and then supporting details. We call these passages $O_1$ (correct passage 1) and $X_1$ (incorrect passage 1). To isolate the effects of specific inconsistencies, we manually revised $O_1$ and $X_1$ to provide the same details but edited to be consistent with the main answer (German or UK) of the passage. Specifically, all passages highlight the crucial role of universities and research institutions in advancing science, but some passages focus on UK institutions while others on German ones. As a result, $O_1$ and $X_1$ followed the Main Answer Inconsistency type.
Then, we used ChatGPT to paraphrase $O_1$ and $X_1$, which simulates Paraphrasing Inconsistency. We manually revised any cases where these paraphrased versions did not provide the same main answer and details as the original passages. As a result, we designed alternative passages $O_2$ and $O_3$ from $O_1$, and $X_2$ and $X_3$ from $X_1$. 

We also designed our comprehension questions based on these inconsistency types. The first question (Q1), focuses on Main Answer Inconsistency by asking about the core answer of the given passage. The second question (Q2) asked about a supporting detail that was semantically similar across passages, but differed due to paraphrasing and was aligned with each passage's main answer. The last question (Q3) asked about a supporting detail that was the same in all passages, albeit paraphrased. 

\subsubsection{Experimental Conditions}\label{exp:condition}

We conducted our experiment as a between-subjects study where each participant received a specific number of passages.
Our study included three conditions: the control condition provided only one passage (Single), and the two different treatment conditions provided two and three passages (Double and Triple, respectively). Furthermore, the conditions were split into subconditions depending on how many correct ($O_{1\sim3}$) and incorrect ($X_{1\sim3}$) passages would be provided to participants, and the order in which correct and incorrect passages appeared on their screens. Based on the possible permutations, there were two subconditions for Single, four for Double, and eight for Triple (total of 14 subconditions, full list in Appendix~\ref{appendix:condition}). 

\subsection{Study Design}


\begin{figure*}[!h]
    \centering
    \includegraphics[width=1.00\textwidth]{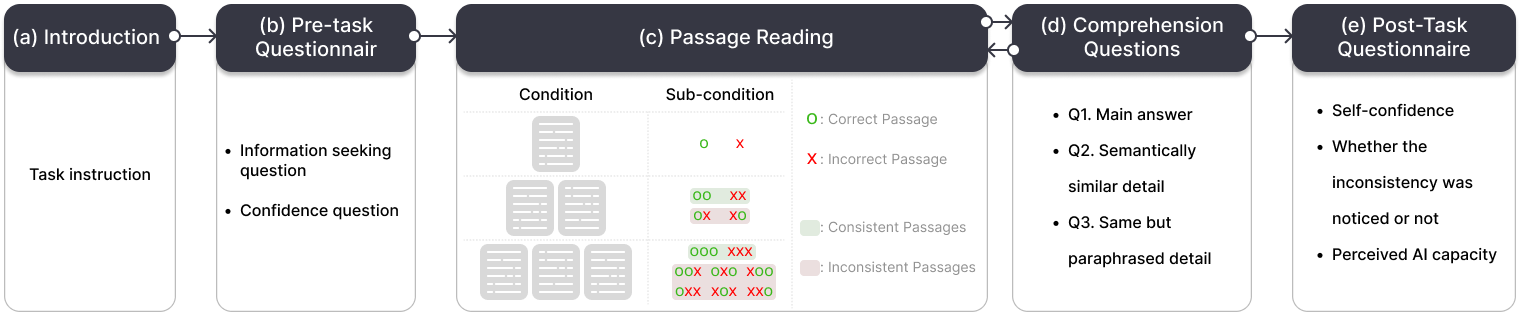}
    \caption{Overall procedure of the experiment. (a) After an introduction to the experiment and task, (b) participants answer the information-seeking question before receiving the passages (i.e., pre-task questionnaire) and rate their confidence. (c) Participants read the AI-generated passage(s) and (d) answer three comprehension questions. (e) Participants respond to post-task questionnaire.}
    \label{fig:procedure}
\end{figure*}

\subsubsection{Study Procedure (Figure~\ref{fig:procedure})}

After accessing our task interface (shown in Appendix~\ref{appendix:interface}), participants were randomly assigned to a subcondition.
(a) Participants received a brief introduction to the study and instructions to follow. If participants were assigned to a condition with multiple passages, they also received an explanation for why they would receive multiple passages. 
(b) Participants were asked to answer the information-seeking question before seeing the AI model's answer(s) and to rate their confidence in their answer on a 7-point Likert scale.
(c, d) Participants then read the given AI-generated passage(s) based on their subcondition, and then answered the three comprehension questions. Participants were not allowed to search for information.
(e) After answering all comprehension questions, participants responded to a post-task questionnaire where they reported if they noticed any inconsistencies and rated their confidence in their answers and their perceived AI capacity on a 7-point Likert scale. We adopted single-item measures since they have been shown to effectively assess perceived AI capacity~\cite{Nourani2019TheEO, papenmeier2022how}. They were also asked to explain the reason for their self-reported ratings.


\subsubsection{Participants and Filtering}

We recruited 252 participants (48.8\% women) through the Prolific crowdsourcing platform\footnote{https://www.prolific.com/} (18 participants for each of the 14 subconditions). 
We only recruited US- and UK-based workers whose first language is English and whose approval rate was higher than 90\%. Participants were paid 0.90£ (about £9/hr payment rate given the 6.7 minute median experiment time). 
As an attention check, we asked participants to explain the passage topic in Figure ~\ref{fig:procedure}-(e). Based on the answers, we filtered out the data of inattentive participants. We ran the study until we recruited 18 participants per subcondition. In total, 285 subjects completed the experiment and we excluded 33 of them.

\subsection{Analysis Methods}


\subsubsection{Variables}
The main independent variable used in our analysis is \textit{inconsistency}, $IC$ (i.e., whether the passages are consistent with each other or not). 
The experimental treatment conditions (i.e., Double or Triple passages) were also considered as key independent variables. 
Although our goal was to see the effect of inconsistencies, various additional factors regarding the passages may change when users are presented with multiple passages, and, as observed in pilot studies, these factors may also influence participants' comprehension and perceived AI capacity. These factors include: the ratio of correct passages among those presented to a participant ($CR_{33}$, $CR_{50}$, $CR_{66}$, $CR_{100}$), and whether the first passage that a participant reads is correct or incorrect ($FP_{first}$) when assuming that they read left-to-right. Thus, to isolate the effect of inconsistencies, we also included these additional factors as independent variables.

For dependent variables, we measured participants' \textit{perceived AI capacity} using their self-reported ratings in the post-task questionnaire: a higher rating meant that participants considered the AI model to have a higher capacity.
Additionally, to measure participants' \textit{comprehension}, we computed the percentage of comprehension questions that they answered correctly ([\textit{number of questions answered correctly}]/[\textit{total number of questions}]\%).

\subsubsection{Statistical Methods}
We start by examining the low-level behavior of participants 
given consistent or inconsistent passages.
We analyzed whether participants' perceived AI capacity (\textbf{RQ1}) and comprehension (\textbf{RQ2}) differed when they consumed consistent or inconsistent passages. 
To answer these questions while avoiding multiple comparison problems and to control for false discovery, we use the interval estimate method~\cite{Dragicevic2016FairSC}. 
We first visualize our data by plotting the means of the dependent variables of interest for each main independent variable (i.e., \textit{consistency}, \textit{the number of passages}) along with the 95\% bootstrap confidence intervals ($R=5000$).
Then, we constructed OLS regression models that predict participants' perceived AI capacity and comprehension scores while controlling for covariates---e.g., participants' demographics (gender, age), whether participants could answer the question correctly before reading the passages ($PQ_{correct}$), and their confidence in their pre-task answers ($PQ_{confidence}$). These models are interpreted through the estimated coefficient values for the independent variables and the bootstrap confidence intervals at 95\%\footnote{Dependent variables were standardized and some independent variables were dummy coded, enabling coefficient of an IV as the change in dependent variable (in terms of standard deviations) resulted from the corresponding treatment.}. 

We also calculated the \textit{``switch fraction''}~\cite{Yin2019UnderstandingTE}: the percentage of participants who followed the AI's answer at the end despite having a different initial answer. 
This measure of switching behavior, as proposed in prior work~\cite{zhang2022you}, serves as a behavioral indicator of users’ reliance on AI, which stems from perceptions of the AI's capacity. Our goal was to understand how self-reported measures of AI capacity manifested in human behavior, providing an additional signal of AI perception.
In cases where participants received multiple passages with inconsistent answers, we consider the AI's answer to be the ``majority'' answer across these passages (e.g., with three passages, if two claimed that Germany had the most Nobel Laureates, then the majority answer was Germany). We did not measure the switch fraction for subconditions without a majority answer (i.e., the same number of correct and incorrect passages). 

Different LLMs have differing capabilities~\cite{open-llm-leaderboard}. Thus, users would likely see different fractions of inconsistent cases depending on the model's accuracy (i.e., the probability of providing correct information). 
To learn how a user group's comprehension and perception of the AI differs with varying model accuracies (\textbf{RQ3}), we performed a bootstrapping analysis with our experiment data. 
To simulate a given model accuracy, we bootstrap from raw data to sample data points for each subcondition such that the occurrences of correct passages across participants, conditions, and passages match the chosen accuracy ($R=500$).
Then, for each accuracy level, we first visualize the bootstrapped data by plotting the mean values for perceived AI capacity and comprehension in each experimental condition.
Additionally, we compare the effect of each condition on perceived AI capacity and comprehension through coefficients from OLS models.
Here, we only considered the experimental treatment conditions as the independent variables as we wanted to see the combined effect of \textit{inconsistency}, \textit{correct ratio}, and \textit{sequence} within each condition. 
We also control covariates by adding the mean of $PQ_{correct}$ and $PQ_{confidence}$ as independent variables.
These models are also interpreted with the estimated coefficient values for the independent variables and 95\% bootstrap confidence intervals. 

\begin{figure*}[!hb]
    \centering
    \includegraphics[width=0.95\textwidth]{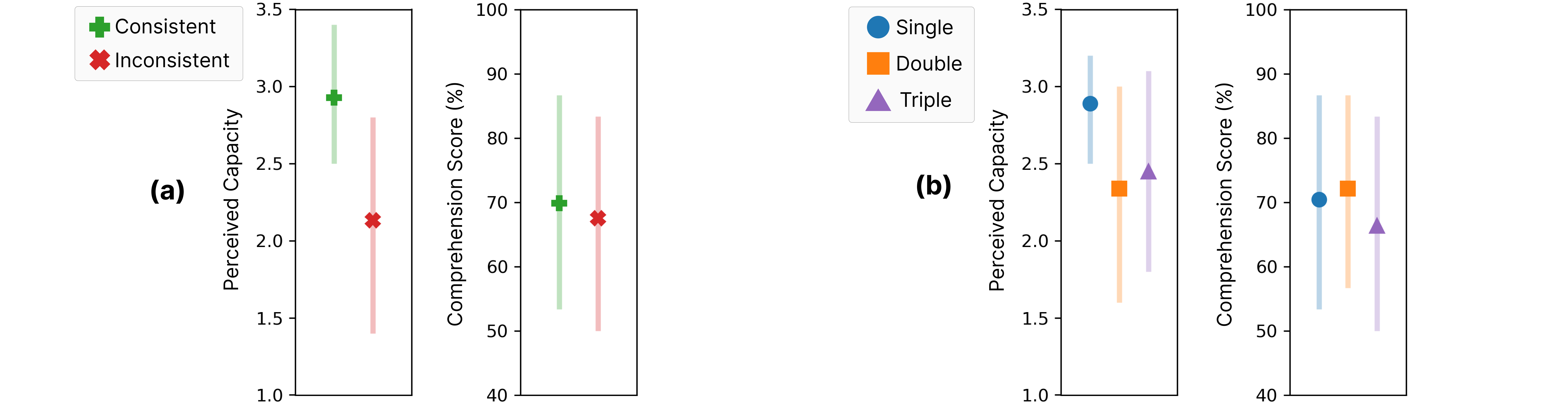}
    \caption{Mean values and 95\% confidence intervals for participants' \textbf{perceived AI capacity} and \textbf{comprehension scores} according to whether they received \textbf{(a)} consistent or inconsistent passages, and \textbf{(b)} their experimental condition (i.e., number of passages).}
    \label{fig:vis-total}
\end{figure*}

\subsubsection{Qualitative Methods}
With a thematic analysis~\cite{Boyatzis1998TransformingQI,Connelly2013GroundedT}, two of the authors qualitatively analyzed responses to the open-ended questions in the post-task questionnaire and categorized them.
This analysis aims to delve deeper into how participants perceived the AI model and how they answered the question with the given passages.

\section{Experimental Results}

In this section, we present our analysis results from the experiment data to answer our research questions. As a sanity check, we first performed the Kruskal-Wallis H Test to examine whether there are any differences across the three conditions regarding participants' $PQ_{correct}$ and $PQ_{confidence}$. We do not find any reliable differences.

\subsection{RQ1: Effects of Inconsistency on Perceived Capacity}\label{finding-rq1}

\begin{figure*}[h!]
    \centering
    \includegraphics[width=0.80\textwidth]{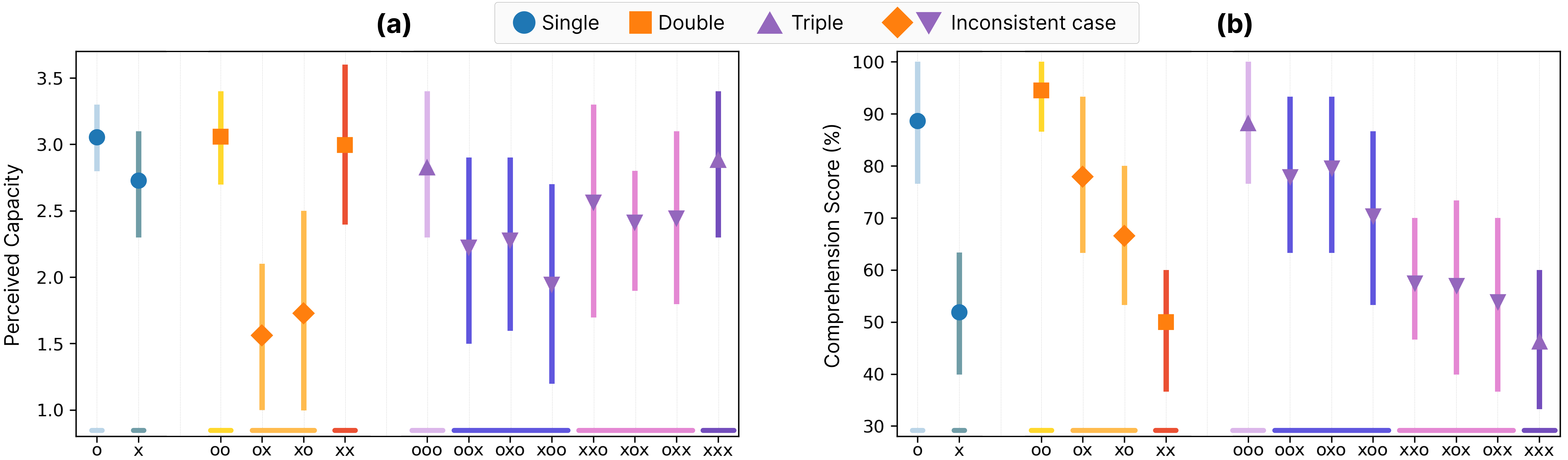}
    \caption{Mean values and 95\% confidence intervals for \textbf{(a)} \textbf{perceived AI capacity} and \textbf{(b)} \textbf{comprehension scores} for each subcondition. The lines above the x-axis labels indicate that these subconditions have the same ratio of correct passages (e.g., [xxo], [xox], and [oxx] provide passages where one-third of them have the correct information).
    }
    \label{fig:vis-subcon}
\end{figure*}

We start by examining the perceived AI capacity of participants who had viewed consistent passages versus those who had inconsistent ones (Figure~\ref{fig:vis-total}-(a)). 
To see the granular effect of each form of inconsistency between passages, we also present a comparison between the perceived capacity of participants in each condition (Figure~\ref{fig:vis-total}-(b)) and subcondition (Figure~\ref{fig:vis-subcon}-(a)).
We find that participants' perceived AI capacity is higher when they received consistent passages compared to those who received inconsistent ones.

We then construct OLS regression models to predict a participant's perceived AI capacity. Our regression results showed that \textbf{inconsistency has a significantly negative effect on perceived capacity} ($\beta$ = -0.1616, 95\% CI = [-0.234, -0.090]), which confirmed the observation that perceived capacity decreases with inconsistent outputs in Figure~\ref{fig:vis-total}-(a). Furthermore, $PQ_{confidence}$ leads to a higher perceived capacity ($\beta$ =  0.1783, 95\% CI = [0.040, 0.316]), indicating that participants with higher confidence have a higher perception of AI capacity.
We further construct two separate regression models for participants who got Double and Triple passages, respectively. The Single condition served as our control condition in these models to compare against the Double and Triple conditions.
To compose an identical set of variables for both models, the independent variable $CR$ (i.e., the ratio of correct passages presented to a participant) was categorized into cases where: the majority answer is correct ($CR_{correct}$); the passages had the same amount of correct and incorrect information ($CR_{tie}$); and the majority is incorrect ($CR_{incorrect}$). 
\textbf{The negative effect of inconsistency on perceived capacity occurred under both conditions} (Double: $\beta$ = -0.1784, 95\% CI = [ -0.256, -0.100]; Triple: $\beta$ = -0.1260, 95\% CI = [-0.225, -0.027]). 
We also found that $PQ_{confidence}$ led to lower perceived capacity in the Triple condition ($\beta$ =  0.3053, 95\% CI = [0.106, 0.50]), suggesting that people with greater confidence perceive AI capabilities more positively. 

From our qualitative analysis on Double-Inconsistent and Triple-Inconsistent cases, we observed that participants noticed conflicting answers but had different opinions about the AI. The majority considered that the AI could not understand information as it was inconsistent, while others thought that the AI could still understand information as its explanations were logical. 
In the former group, D31 (Double-Inconsistent, participant \#31) stated that \textit{``it gave a different answer for each passage. How do you know which one is accurate? You can't trust it to give you a real answer.''} T30 (Triple-Inconsistent, participant \#30) indicated that \textit{``the AI model had two different answers to the question, so I am not sure it had the best grasp on things''}.
In the latter group, D35 mentioned that \textit{``the lack of conclusive results makes it difficult to say for certain, but the reasoning behind it seemed sound.''}
Interestingly, some participants attributed ``human'' characteristics to the model by saying that it can ``get confused'' while others thought the AI assistant ``is programmed with a database or information''.


\begin{table}[!t]
\centering
    \begin{tabular}{cccl} 
        \toprule
        \textbf{$PQ_{correct}$} & \textbf{Condition} & \textbf{Q1 Correct} & \textbf{Switch Fraction} \\
        \midrule
        x & o & o & 0.9167 (11/12)\\
        o & x & x & 1.0000 (3/3)\\
        x & oo & o & 1.0000 (14/14) \\
        o & xx & x & 1.0000 (7/7)\\
        x & ooo & o & 0.9167 (11/12)\\
        o & xxx & x & 0.8889 (8/9)\\
        \midrule
        x & oox, oxo, xoo & o & 0.7105 (27/38)\\
        o & xxo, xox, oxx & x & 0.6875 (11/16)\\
        \bottomrule
    \end{tabular}
\caption{The correctness of participants' final answer to Q1 and their switch fraction depending on their subcondition and whether their pre-task answer to Q1 was correct or not.}
\Description{The headers are $PQ_{correct}$, Passage condition, Correctness of answer to Q1, and Switch Fraction.}
\label{tab:switch-fraction}
\end{table}

To understand how self-reported measures of AI capacity manifested in observed behaviors, we examined whether participants followed the information provided by the AI. Specifically, we calculated the switch fraction, which represents the proportion of participants who changed their initial answer to align with the AI's response. The results for each subcondition are in Table~\ref{tab:switch-fraction}.
Out of all the participants that received consistent passages and had a pre-task answer different from the model's answer, only three participants did not follow the model's majority answer---revealing that participants tended to overrely on the model when it is consistent.
On the contrary, when participants received inconsistent passages, \textbf{a higher proportion of those who had a main answer that differed from the model’s answer stuck to their answer even after seeing the model's answer.}
Our qualitative analysis of the comments of these participants revealed that, once they noticed the inconsistency, they chose the passage that seemed \textit{``more logical''} (T16), \textit{``fluent''} (T14), or aligned with their prior knowledge (T13).
These findings imply that inconsistent passages could encourage participants to critically examine the output to reach a decision, rather than simply following the AI model's answer.

\subsection{RQ2: Effects of Inconsistency on Comprehension}\label{finding-rq2}

The comprehension of the participants with consistent passages was similar to that of those who received inconsistent passages (Figure~\ref{fig:vis-total}-(a)). 
As in Figure~\ref{fig:vis-subcon}-(b), which compares the comprehension scores of the participants between the subconditions, the participants had higher comprehension as $CR$ increases and/or when the first passage was correct.

Since it is challenging to observe the direct effect of inconsistencies in comprehension scores due to multiple factors that could moderate the effect, we performed a regression analysis to examine whether participants' comprehension is influenced by inconsistency when including multiple independent variables.
Our results on the mean comprehension score of the total questions showed that \textbf{the comprehension of the participants increased significantly when there was inconsistency} ($\beta$ = 0.1329, 95\% CI = [0.065, 0.201]). 
$CR_{100}$ and $CR_{66}$ also lead to a higher comprehension score ($CR_{100}$: $\beta$ = 0.3681, 95\% CI = [0.251, 0.486], $CR_{66}$: $\beta$ = 0.1349, 95\% CI = [0.063, 0.207]). 
We constructed two separate models to analyze participant comprehension under the Double and Triple conditions, categorizing $CR$ into three groups, as we did in the analysis of perceived capacity. 
Different effects emerged in the two separate models.
\textbf{Although inconsistency has a significantly positive effect on comprehension in the Double condition} ($\beta$ = 0.1080, 95\% CI = [0.034, 0.182]), there is \textbf{no significant effect on comprehension in the Triple condition.}
$CR_{correct}$ leads to a higher comprehension score under both conditions ($CR_{correct}$ in Double: $\beta$ = 0.3569, 95\% CI = [0.150, 0.564], $CR_{correct}$ in Triple: $\beta$ = 0.2323, 95\% CI = [0.135, 0.329]). 
Our qualitative analysis of responses from Triple-inconsistent participants shows that they seemed to follow the \textbf{majority} answer. T89 mentioned that ``\textit{I was confident because Germany came out twice out of the three passages.}''

\begin{figure*}[hb!]
    \centering
    \includegraphics[width=0.95\textwidth]{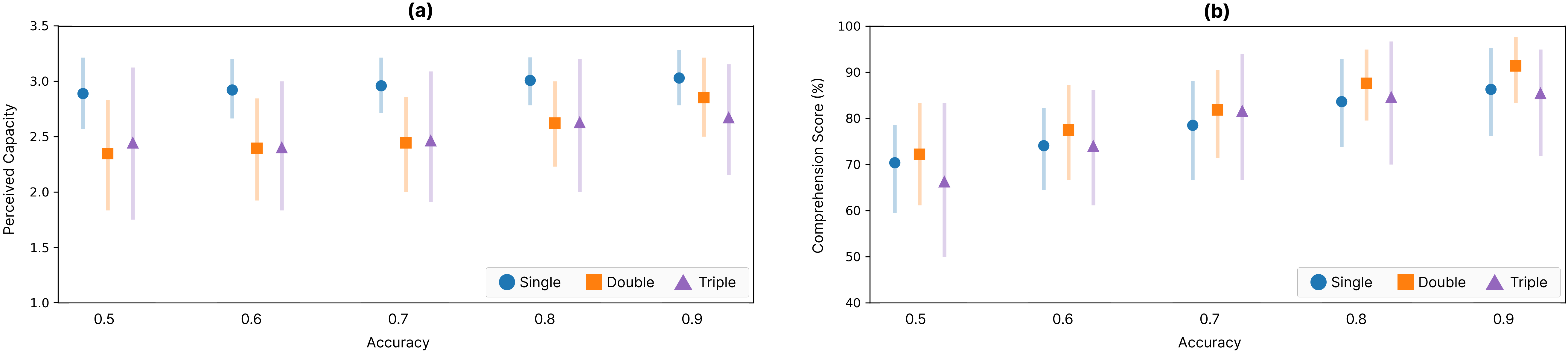}
    \caption{\textbf{(a)} Comparison of mean \textbf{perceived AI capacity} and 95\% intervals across different levels of AI model accuracy. \textbf{(b)} Same visualization for \textbf{comprehension scores}.}
    \label{fig:vis-acc}
\end{figure*}

Our comprehension questions focused on different parts of the passages, each embodying a different type of inconsistency. Analyzing comprehension scores on each individual question can provide deeper insight into how inconsistency in the main answer affects detailed behavior outcomes. 
\textbf{Inconsistency has a significantly positive effect on comprehension in Q1} (IC: $\beta$ = 0.2364, 95\% CI = [0.140, 0.333]) and $PQ_{correct}$ ($\beta$ = 0.1440, 95\% CI = [0.053, 0.234]). 
$CR_{33}$, $CR_{66}$, $CR_{100}$, and $FP_{first}$ also have a significant effect on the comprehension score of Q1 ($CR_{33}$: $\beta$ = -0.1897, 95\% CI = [-0.287, -0.092], $CR_{66}$: $\beta$ = 0.3583, 95\% CI = [0.257, 0.460], $CR_{100}$: $\beta$ = 0.7011, 95\% CI = [0.535, 0.867], $FP_{first}$: $\beta$ = -0.2519, 95\% CI = [-0.364, -0.140]). These results indicate that according to the correctness of the majority answer, participants' comprehension is also affected and, if the incorrect answer is shown first, comprehension decreases. 
The positive effect of inconsistency on the comprehension of Q1 is in line with our qualitative findings. Participants mentioned that, once they noticed inconsistencies between passages, they \textit{``reconsidered''} the information (T90). T30 also stated that \textit{``the differences forced me to stop and consider the information I was seeing in each passage, and how much of it lined up/correlated between different passages''}
Additionally, we observed that inconsistencies encouraged some participants to rely on their pre-task answer or prior knowledge. D1 indicated that \textit{``one of the two different responses from the AI agreed with [my prior answer] so that may suggest it could be correct.''}

\textbf{Inconsistency also had a significantly positive effect on comprehension in Q2}, which asked about information paraphrased across the passages ($\beta$ =0.1541, 95\% CI = [0.015, 0.294]). 
This can imply that, as participants noticed an inconsistency in passages, this encouraged them to read passages more carefully and led them to understand information that was explained differently in the passages.
Furthermore, $CR_{100}$, and $PQ_{confidence}$ also have a significant effect on the Q2 comprehension score ($CR_{100}$: $\beta$ = 0.3051, 95\% CI = [0.065, 0.545], $PQ_{confidence}$ : $\beta$ = -0.2740, 95\% CI = [-0.541, -0.007]).
Qualitative findings also show that participants read and thought carefully when they identified differences between answers. D30 said \textit{``Having different answers to the question made me think more about what the text was saying.''} 

Inconsistency had no significant effect on comprehension in Q3, which asked about information that was the same across passages.


\subsection{RQ3: How the Effects of the Number of Passages Change when Model Accuracy Changes}\label{finding-rq3}

Through bootstrapping, we explored how the treatment's impact on perceived capacity and comprehension varies according to accuracy levels, which are captured by different fractions of inconsistent cases that a user group receives.

\subsubsection{RQ3-1: How Effect of Passage Number on Perceived Capacity Changes}\label{finding-rq4-1}

First, we examined how the number of passages affects users' perceived capacity depending on the model's accuracy or capability to provide correct information.
In figure~\ref{fig:vis-acc}-(a) showing the perceived capacity of participants in all conditions at all accuracy levels, participants in the Single condition reported the highest perceived capacity compared to the Double and Triple conditions.
We constructed OLS regression models to predict a participant's perceived capacity with bootstrapped data for each accuracy level. 
Figure~\ref{fig:vis-pattern}-(a) compares coefficients of treatment across accuracy levels ranging from 50\% to 98\% with a step of 2\%.
We find that, \textbf{regardless of accuracy, both multiple passage conditions have a negative effect on perceived capacity}. 
As the model accuracy increases, the negative effect of the Double condition on perceived capacity decreases, which might be caused by the decreasing number of inconsistent Double passages. 
However, we also observe that there is not as much coefficient change in the Triple condition compared to the Double condition. 
This implies that the Triple condition may have a negative effect on perceived capacity across the accuracy levels, which could lead to the potential risk that users do not rely on the AI even when the AI is very capable.

\subsubsection{RQ3-2: How Effect of Passage Number on Comprehension Changes}\label{finding-rq4-2}

\begin{figure*}[h!]
    \centering
    \includegraphics[width=0.95\textwidth]{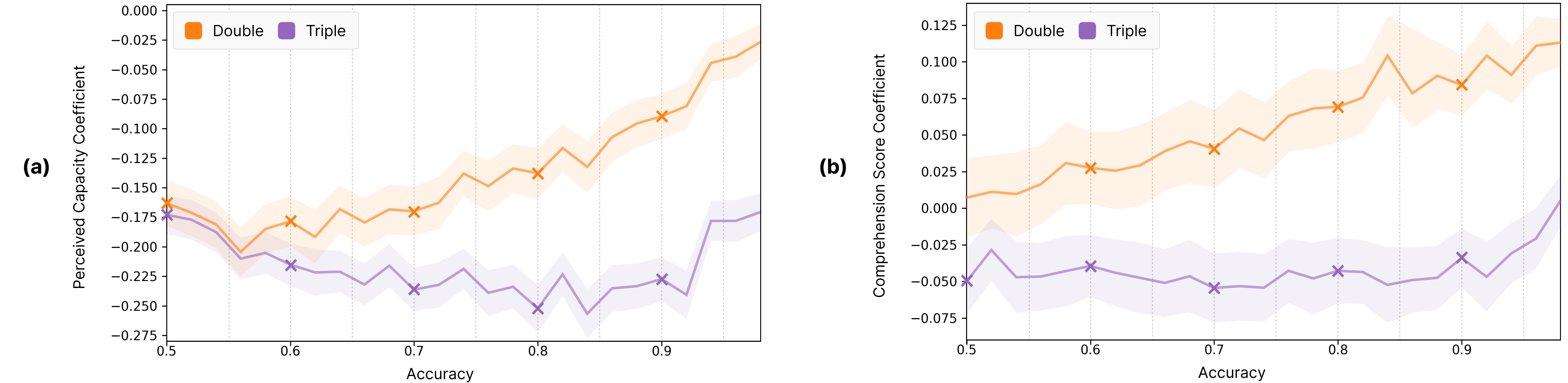}
    \caption{\textbf{(a)} Coefficient of the Double and Triple conditions on \textbf{perceived AI capacity} across accuracy levels for the AI model (50\%-98\%, step of 2\%). The shaded region represents the 95\% confidence intervals. Each 10\% step of accuracy marked with a cross represents that the coefficient was statistically significant (p<.05). \textbf{(b)} Same visualization for \textbf{comprehension scores}.}
    \label{fig:vis-pattern}
\end{figure*}

We next examine how the number of passages affects participants' comprehension of the information when the model accuracy changes.
Figure~\ref{fig:vis-acc}-(b) shows the comprehension scores for each condition at all accuracy levels.
In lower accuracy, the comprehension score of participants under different conditions appears to be similar to each other. 
As accuracy increases, the Double condition has a slightly higher mean comprehension score compared to the other two conditions, and there are bigger gaps between the lower bounds of the conditions. Additionally, in all accuracy levels, variance in the Triple condition is the largest.
Figure~\ref{fig:vis-pattern}-(b) compares coefficients of treatment across accuracy levels ranging from 50\% to 98\% with a step of 2\%, which can show changes in treatment's impact on comprehension according to accuracy changes.
\textbf{In all accuracy levels, we find that the Double condition has a positive effect on comprehension, whereas the Triple condition has a negative effect.} 
Moreover, it appears that as model accuracy increases, the Double condition's positive effect on comprehension score increases, while there is less change in the coefficient of the Triple condition. 
Specifically, as accuracy is close to 100\%, the coefficient of the Triple condition goes from negative to zero, which means that it becomes non-detrimental.

\section{Discussion}

Here, we interpret the results from our study and propose implications for the design of LLM-powered systems based on our findings.

\subsection{Interpretation of Results}
While prior work regarded inconsistency as a problem, our study showed that revealing inconsistencies can be beneficial for users when consuming AI-generated outputs. 
Our study found that inconsistency lowers the perceived AI capacity of the participants, as they considered that AI could ``get confused'' or contain ``conflicting information'' (Section ~\ref{finding-rq1}).
We also observed that inconsistencies had a positive effect on participants' comprehension of details within the generated information (Section~\ref{finding-rq2}).
Based on the theory of desirable difficulties~\cite{bjork2011making}, we posit that this finding can be attributed to how inconsistencies in AI outputs, which participants perceived as challenging yet manageable, may have encouraged them to engage more intensively with the content and read more thoroughly.
The effect of inconsistencies in moderating trust in the AI also led participants to be more open to considering their own thoughts, rather than simply following AI, which is aligned with findings from prior literature~\cite{Yin2019UnderstandingTE} that showed how performance indicators can affects how people revise their decisions based on AI predictions. In other words, inconsistency had a positive effect on comprehension score regarding the main answer (Section~\ref{finding-rq2}) and more participants did not follow the AI's majority answer when reading inconsistent passages (Section~\ref{finding-rq1}).

In situations when AI accuracy is above 50\%, we expected that the Triple condition would have a positive effect on comprehension, as it provides more passages and increases the likelihood that participants receive at least one passage with correct information. 
However, the findings of Section~\ref{finding-rq3} showed that the three-passage condition consistently hindered comprehension across all accuracy levels.
We observed that this may be due to the way that most participants relied on simple majority voting as a heuristic---similar to software fault tolerance~\cite{mackie1987systematic}---and ignored the correct passage when it was in the minority.
Then, considering how participants resorted to majority voting, it can be assumed that the triple condition would have a positive effect on comprehension as accuracy increases, since most participants will receive passages where the majority is correct.
However, we observed that the triple condition still had a negative impact on comprehension, potentially due to the increased cognitive load from reading more passages, as per the Cognitive Load Theory~\cite{karr2010more}. 
Due to this load, participants may have skimmed the passages, preventing them from critically comparing the passages and leading them to overlook detailed information. 

\subsection{Design Implications}
When designing an LLM-powered system for information consumption, we suggest that the system should tailor the amount of information generated (e.g., generated answer, proposed action) based on the model's performance level.
When the model's performance is notably high (i.e., higher than 90\%), the prevalent system design of providing a single output would be sufficient, as it is unlikely for the model to produce incorrect outputs and it can be reasonable for the user to rely on the LLM---showing more outputs introduces the unnecessary risk of users learning incorrect information.
When employing models with lower accuracy, we suggest that systems should transition to designs that provide two outputs, as our study showed that this can prevent overreliance and encourage more critical reading, while maintaining cognitive load at a manageable level~\cite{1331674}.
As model accuracy depends on the information domain or user context, we propose that developers can construct specific benchmarks to assess the models' accuracy in domains of interest to then decide on how the system's design should be adapted.
While our study explored this transition from single- to multi-output designs in the context of information-seeking, this transition can also be effective for tasks where presenting users with multiple perspectives can promote more nuanced understanding, such as healthcare decision-making or news consumption.
Finally, as revealed in our qualitative results, this transition can also be beneficial when it is important to trigger users' intrinsic motivation to find more information---e.g., encourage students to look for more information to resolve an inconsistency between system outputs~\cite{bjork2011making}.

While we advocate for providing two outputs to users, we acknowledge the potential trade-off of presenting multiple outputs as they require more time and cognitive effort to compare and critically analyze. This increased processing load, essential for gaining deeper insights, might be a deterrent in real-life scenarios where users seek quick information or face time constraints. Therefore, the challenge lies in balancing the depth of content with the user's capacity and willingness to engage with it.
To address this, we recommend providing visual support to facilitate users' processing of multiple outputs.
Inspired by prior work on supporting sensemaking of text~\cite{texsketch}, we recommend gradually showing automated highlights of the differences between outputs by calculating the similarity between sentences and enabling users to annotate their interpretation on them.
To facilitate sensemaking of these differences, identified inconsistencies could also be labeled with their types, as identified in Section~\ref{sec:preliminary_results}.

In our study, we limited our comparison to double and triple as multiple passages due to technical-side and user-side concerns when employing a larger number of passages in real-world scenarios.
Although the current generation speed of models may be too slow, advancements in model efficiency could allow real-time applications to produce large numbers of outputs for users. 
Future work can explore novel ways to help users assess the reliability of LLMs by comparing large samples. 
Using fact verification methods~\cite{Min2023FActScoreFA} to automatically distinguish between passages, future systems could display a variety of unique passages to avoid redundancy.
Furthermore, by automatically identifying passages that share answers, system builders could calculate ``confidence scores'' for specific answers (i.e., percentage of outputs that share the answer)---as suggested in Leiser et al.~\cite{leiser2023chatgpt}. Then, the system can present users with passages for the top-2 answers---leveraging the benefits of double passages while also providing reliability indicators. 

While we discussed design implications focusing on LLMs, these guides can be broadly applied to the design and development of generative AI systems. We suggest, however, that developers should take into account how the output of each generative AI is consumed (e.g., time and effort required). For instance, users can find it easier to process and compare multiple image outputs, in contrast to text, speech or video outputs.

\subsection{Intuitive Intervention to Build User's Mental Models of LLMs}
To prevent users from overrelying on LLM outputs, we propose an intuitive intervention of showing inconsistencies in the outputs. 
First, by seeing inconsistencies, users can more intuitively realize the nondeterministic nature of LLMs. This awareness can encourage users to adopt a critical perspective when interacting with these models, which can be crucial as LLMs produce more convincing outputs than previous AI models.
This can also lead users to become curious about why the LLM was inconsistent, which can encourage them to understand the mechanics behind these model's training and inference.
Encouraging this understanding can be critical, as some of our participants misunderstood why the inconsistencies were produced.
For example, some thought that the model \textit{retrieved incorrect information} or became \textit{confused as it read conflicting information}, rather than understanding the actual word-by-word probabilistic generation process. 
Thus, by presenting inconsistencies and the mechanisms behind them in a more transparent and explainable manner, systems could help users to better grasp the mechanics and limitations of these models, fostering more informed and critical consumption of AI-generated content.
This method of displaying inconsistencies could serve as a more generalizable approach to representing uncertainty, a task that is challenging in standard chat interfaces~\cite{Marsi2007ExpressingUW}.

\subsection{Limitations \& Future Work}
While it is unclear whether our findings would also hold for other passage types, the insights gained may still apply to similar types of text that prioritize factual accuracy and the presentation of answers with supporting details. 
The passage types influence the user's cognitive load, which can impact their ability to comprehend multiple passages.
For example, passages with logical or mathematical reasoning might incur a higher cognitive load, causing users to struggle even when reading only two passages.
Second, our study design was limited to a single turn of interaction with AI. 
This leaves room for future research on how the experience of reading a passage might affect subsequent usage of AI. 
Third, it is not clear how our study results were influenced by the pre-task question, which was used to control the prior knowledge effect.
We also recognize that the order of questions on perceived AI capacity and inconsistencies could have affected outcomes, although these effects would likely be consistent across conditions.
Finally, since we only focused on two types of inconsistencies, studying other types would provide more insight into how inconsistencies impact user behavior.
\section{Conclusion}

This work examines the impact of inconsistencies in LLM's generated outputs on how users interact with LLMs in an information-seeking context.
Through a qualitative analysis, we first identified five types of inconsistencies that can occur between the LLM-generated output. We conducted an online experiment (N=252) to understand the effect of providing users with multiple outputs that can contain inconsistencies.
Our findings suggest that inconsistent output negatively impacted participants' perceived AI capacity while positively influencing their comprehension of the generated information, especially for participants who received double passages. 
Based on these findings, we suggest that, instead of viewing inconsistencies as a limitation, future systems could leverage inconsistencies to mitigate users' overreliance on AI and encourage critical examination of generated outputs.

\begin{acks}
This work was supported by the Institute of Information \& Communications Technology Planning \& Evaluation (IITP) grant funded by the Korean government (MSIT) (No.2021-0-01347, Video Interaction Technologies Using Object-Oriented Video Modeling).
\end{acks}

\bibliographystyle{ACM-Reference-Format}
\bibliography{references}
\newpage
\onecolumn
\appendix

\section{Inconsistency Types and Examples}
\label{appendix:incon-type}
\begin{table*}[h!]
\begin{tabular}{p{2.5cm}p{4cm}p{5.0cm}p{5.0cm}}
\toprule
\textbf{Type} & \textbf{Passage A} &\textbf{Passage B} &\textbf{Description} \\ \midrule
\textbf{Main Answer Inconsistency}    & As of my last knowledge update in September 2021, \textit{China} was the large country that spent the most on international tourism;  & As of my last knowledge update in September 2021, the \textit{United States} was the large country that spends the most on international tourism. & Passage A explains the answer is China, and Passage B responds the answer is United States on the same question. \\ \hline
\textbf{Detail Volume Inconsistency} & [Main answer A,  supporting detail a and b]   & [Main answer A,  supporting detail a and b] \textit{Additionally, ethical considerations and practical limitations often make raising a chimpanzee as a human child an impractical and controversial endeavor.} & Compared to Passage A, Passage B explains additional information related to the ethical considerations and practical limitations of raising a chimpanzee as a human child.\\ \hline
\textbf{Detail Content Inconsistency}& Coffee has become deeply ingrained in American culture, serving as a ubiquitous morning ritual and a social beverage & Coffee has a deep roots in American history, dating back to colonial period, and it became a symbol of the American Revolution as an alternative British tea. & Passage A emphasizes contemporary cultural aspects of coffee in American life, while Passage B introduces historical context, creating a difference in the level of detail and focus on the historical roots of coffee.\\\hline
\textbf{Detail Expansion Inconsistency} & In fact, \textit{gluten-free products often lack certain nutrients}   & In fact, gluten-free products often lack the same nutritional content and \textit{may contain added sugars and fats to compensate for taste and texture}. & Passages A and B both say that gluten-free products lack certain nutrients, but Passage B additionally explains what kind of nutrients they are. \\ \hline
\textbf{Paraphrasing Inconsistency} & Lastly, the fast-paced American lifestyle often necessitates a caffeine boost, further driving the demand for coffee.& The demands of a fast-paced lifestyle, which all contribute to the high coffee consumption in the United States. & Passages A and B both convey the same information with different styles.\\ \bottomrule
\end{tabular}
\caption{The four types of inconsistency that were identified between pairs of outputs generated by LLM in our Preliminary Study. Each type of inconsistency is illustrated with example passages.}
\Description{The headers are type, passage A, passage B, and description of the type.}
\end{table*}

\section{Interface}
\label{appendix:interface}
\begin{figure*}[ht!]
    \centering
    \includegraphics[width=1.00\textwidth]{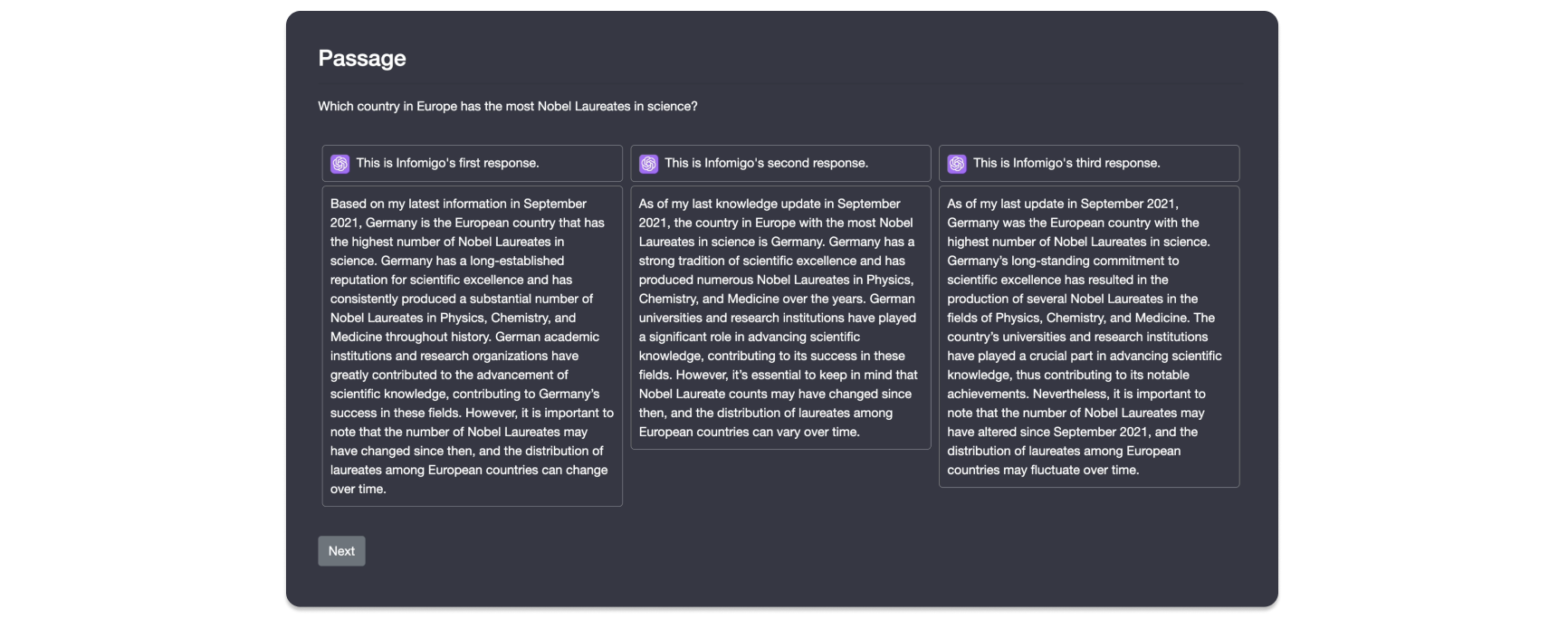}
    \caption{The interface employed during the experiment showing the procedure step where participants are presented with passages generated by the AI assistant. The example demonstrates the triple condition where participants receive three passages.}
    \label{fig:interface}
\end{figure*}

\section{Details in Conditions}
\label{appendix:condition}
\begin{table*}[h!]
\begin{tabular}{ccc}
\toprule
\textbf{Condition} & \textbf{subconditions} & \textbf{Passage participants received} \\ \midrule
\multirow{2}{0.05\textwidth}{Single}   & O & $O_k$ ($k\in\{1, 2, 3\}$)\\
  & X & $X_k$ ($k\in\{1, 2, 3\}$)  \\\midrule
\multirow{4}{0.05\textwidth}{Double} 
    & OO & $O_{k}O_{l}$ ($k \neq l$, $k, l\in\{1, 2, 3\}$)   \\
    & OX & $O_{k}X_{l}$ ($k \neq l$, $k, l\in\{1, 2, 3\}$)  \\
    & XO & $X_{k}O_{l}$ ($k \neq l$, $k, l\in\{1, 2, 3\}$)  \\
    & XX & $X_{k}X_{l}$ ($k \neq l$, $k, l\in\{1, 2, 3\}$)  \\\midrule
\multirow{8}{0.05\textwidth}{Triple} 
& OOO & $O_{k}O_{l}O_{m}$ ($k \neq l, l \neq m, k \neq m$, $k, l, m\in\{1, 2, 3\}$)   \\
    & OOX & $O_{k}O_{l}X_{m}$ ($k \neq l, l \neq m, k \neq m$, $k, l, m\in\{1, 2, 3\}$)   \\
    & OXO & $O_{k}X_{l}O_{m}$ ($k \neq l, l \neq m, k \neq m$, $k, l, m\in\{1, 2, 3\}$)  \\
    & XOO & $X_{k}O_{l}O_{m}$ ($k \neq l, l \neq m, k \neq m$, $k, l, m\in\{1, 2, 3\}$)\\
    & OXX & $O_{k}X_{l}X_{m}$ ($k \neq l, l \neq m, k \neq m$, $k, l, m\in\{1, 2, 3\}$)\\
    & XOX & $X_{k}O_{l}X_{m}$ ($k \neq l, l \neq m, k \neq m$, $k, l, m\in\{1, 2, 3\}$)\\
    & XXO & $X_{k}X_{l}O_{m}$ ($k \neq l, l \neq m, k \neq m$, $k, l, m\in\{1, 2, 3\}$)\\
    & XXX & $X_{k}X_{l}X_{m}$ ($k \neq l, l \neq m, k \neq m$, $k, l, m\in\{1, 2, 3\}$)\\
\bottomrule
\end{tabular}
\caption{Condition table that includes subconditions that are allocated to each condition and type of passage that participant in each subcondition received.}
\Description{The headers are condition, subconditions, and passage participants received.}
\label{tab:condition}
\end{table*}









\end{document}